\documentclass[aps,prd,draft,showpacs]{revtex4}
\usepackage[final]{graphicx}
\begin{document}

\newcommand{\be}{\begin{equation}}
\newcommand{\ee}{\end{equation}}
\newcommand{\ben}{\begin{eqnarray}}
\newcommand{\een}{\end{eqnarray}}

\newcommand{\la}{{\lambda}}
\newcommand{\Om}{{\Omega}}
\newcommand{\ta}{{\tilde a}}
\newcommand{\bg}{{\bar g}}
\newcommand{\bh}{{\bar h}}
\newcommand{\si}{{\sigma}}
\newcommand{\C}{{\cal C}}
\newcommand{\D}{{\cal D}}
\newcommand{\cA}{{\cal A}}
\newcommand{\cT}{{\cal T}}
\newcommand{\cO}{{\cal O}}
\newcommand{\eeo}{\cO ({1 \over E})}
\newcommand{\G}{{\cal G}}
\newcommand{\cL}{{\cal L}}
\newcommand{\T}{{\cal T}}
\newcommand{\M}{{\cal M}}

\newcommand{\p}{\partial}
\newcommand{\na}{\nabla}
\newcommand{\ssum}{\sum\limits_{i = 1}^3}
\newcommand{\dssum}{\sum\limits_{i = 1}^2}
\newcommand{\tal}{{\tilde \alpha}}

\newcommand{\tp}{{\tilde \phi}}
\newcommand{\tPhi}{\tilde \Phi}
\newcommand{\tpsi}{\tilde \psi}
\newcommand{\tim}{{\tilde \mu}}
\newcommand{\tr}{{\tilde \rho}}
\newcommand{\tir}{{\tilde r}}
\newcommand{\rp}{r_{+}}
\newcommand{\hr}{{\hat r}}
\newcommand{\rv}{{r_{v}}}
\newcommand{\dr}{{d \over d \hr}}
\newcommand{\dR}{{d \over d R}}

\newcommand{\hhf}{{\hat \phi}}
\newcommand{\hhM}{{\hat M}}
\newcommand{\hhQ}{{\hat Q}}
\newcommand{\hht}{{\hat t}}
\newcommand{\hhr}{{\hat r}}
\newcommand{\hhS}{{\hat \Sigma}}
\newcommand{\hhD}{{\hat \Delta}}
\newcommand{\hhm}{{\hat \mu}}
\newcommand{\hro}{{\hat \rho}}
\newcommand{\hhz}{{\hat z}}

\newcommand{\tD}{{\tilde D}}
\newcommand{\tB}{{\tilde B}}
\newcommand{\tV}{{\tilde V}}
\newcommand{\hT}{\hat T}
\newcommand{\tF}{\tilde F}
\newcommand{\tT}{\tilde T}
\newcommand{\hC}{\hat C}
\newcommand{\ep}{\epsilon}
\newcommand{\bep}{\bar \epsilon}
\newcommand{\ppp}{\varphi}
\newcommand{\Ga}{\Gamma}
\newcommand{\ga}{\gamma}
\newcommand{\hth}{\hat \theta}
\title{Thick Domain Walls and Charged Dilaton Black Holes}

\author{Rafa{\l} Moderski}

\affiliation{N.Copernicus Astronomical Center \protect \\
Polish Academy of Sciences, \protect \\
00-716 Warsaw, Bartycka 18, Poland \protect \\
moderski@camk.edu.pl}

\author{Marek Rogatko}

\affiliation{Institute of Physics \protect \\
Maria Curie-Sklodowska University \protect \\
20-031 Lublin, pl.Marii Curie-Sklodowskiej 1, Poland \protect \\
rogat@tytan.umcs.lublin.pl \protect \\
 rogat@kft.umcs.lublin.pl}

\date{\today}
\bigskip
\begin{abstract}
We study a black hole domain wall system in dilaton gravity which is
the low-energy limit of the superstring theory. We solve numerically
equations of motion for real self-interacting scalar field and
justify the existence of static axisymmetric  field configuration 
representing the thick domain wall in the background of a charged
dilaton black hole. It was also confirmed that the extreme dilaton black 
hole always expelled the domain wall.
\end{abstract}
\maketitle
\baselineskip=18pt
\section{Introduction}
It is believed that the early universe undergoes a series of vacuum
phase transitions which led to several types of topological
defects \cite{vil}. Topological defects arising in the early universe could
envisage the high-energy phenomena that were beyond the range of our accelerators.
The studies
of topologically nontrivial field configurations in the 
background of a black hole attract great interest in recent years.
In Refs.\cite{abh} it was shown both analitycally and numerically that
an Abelian-Higgs vortex could act as a long hair for the
Schwarzschild and Reissner-Nordstr\"om (RN) black hole solution.
\par
The problem of dilaton black hole cosmic string system was
considered in Refs.\cite{dil,san00}.
It was revealed that the horizon of a charged dilaton black hole 
could support a long-range fields of the Nielsen-Olesen type, which one
could consider as black hole hair. If the dilaton black hole approached 
to extremality one could show that the vortex was always expelled from it.
\par
De Villier {\it et al.} \cite{vil98} studied the dynamics of the scattering
and capturing process of an infinitely thin cosmic string in the background of
Schwarzschild black hole spacetime.\\
The domain wall in the black hole background was considered by Christiensen
{\it et al.} \cite{chr98}, showing that there exists family of infinitely thin walls
intersecting black hole event horizon. 
\par
In Ref.\cite{hig00} the stability of a Nambu-Goto membrane at the equatorial
plane in the background of RN-de Sitter spacetime was studied. It was shown
that  a membrane intersecting charged black hole is unstable and the positive
cosmological constant strengthens this instability.\\
The gravitational interaction of a
thick domain wall in the Schwarzschild black hole background was studied in 
Ref.\cite{mor00}.
Bonjour {\it et al.} \cite{bon99} have investigated the spacetime of thick gravitating
domain wall with local  planar symmetry and reflection symmetry around
the wall's core. They revealed that the domain wall spacetime has a
cosmological horizon as in the de Sitter case.\\
Recently, the interaction of black holes and extended objects attracts the attention
in the context of superstring/M-theory and the brane world scenario \cite{hor96}.
It is argued that in this scenario black holes on the gravitating membrane 
are realized as {\it black cigars} in the bulk spacetime intersecting membrane.
In Ref.\cite{emp01} the problem of black hole on a topological domain wall
(including the gravitational back-reaction) was considered. In \cite{rog01}
the behaviour of the domain wall in the spacetime of a dilaton black hole
was analyzed and it was shown analytically that the extreme dilaton black 
hole always expelled the domain wall.
\par
In our paper we shall 
provide some continuity with the previous work \cite{rog01} and we shall
consider the interaction between a domain wall and a
charged dilaton black hole taking into account
the thickness of the domain wall and the potential
of the scalar field of $\varphi^4$ and sine-Gordon forms.
\par
The outline of the paper is as follows. Sec.II is devoted to the general 
analytic considerations of the domain wall in the spacetime of the
charged dilaton black hole. In Sec.III we presented the numerical analysis of the
equation of motion for two cases of potentials with discrete set
of degenerate minima, i.e., the $\varphi^4$ and sine-Gordon potentials.
In Sec.IV we finish with the general summarizing of our work.

\section{The basic equations of the problem}
In this section we shall consider a static thick domain wall in 
the background of a charged dilaton black hole. This black hole is
a static spherically symmetric solution of Einstein-dilaton gravity being 
the low-energy limit of the superstring theory. In our considerations
we assume that the domain wall is constructed by means of a self-interacting
scalar field in the considered background.
The metric of a charged dilaton black hole may be written as
\cite{gar91}
\be
ds^2 = - \left ( 1 - {2 M \over r} \right ) dt^2 +
{ d r^2 \over \left ( 1 - {2 M \over r} \right ) } + r \left ( 
r - {Q^2
\over M} \right ) (d \theta^2
+ \sin^2 \theta  d \ppp^2),
\ee
where we define $r_{+} = 2M$ and $r_{-} = {Q^2 \over M}$ which are
related to the mass $M$ and charge $Q$ by the relation $Q^2 = {r_{+}
r_{-} \over 2} e^{2 \phi_{0}}$. The charge of the dilaton black hole
$Q$, couples to the field $F_{\alpha \beta}$.  The dilaton field is
given by $e^{2\phi} = \left ( 1 - {r_{-} \over r} \right )
e^{-2\phi_{0}}$, where $\phi_{0}$ is the dilaton's value at $r
\rightarrow \infty$. The event horizon is located at $r = r_{+}$. For
$r = r_{-}$ is another singularity, one can however ignore it because
$r_{-} < r_{+}$. The extremal black hole occurs when $r_{-} = r_{+}$,
when $Q^2 = 2M^2 e^{2\phi_{0}}$.
\par

We consider a general matter Lagrangian with real Higgs field and the
symmetry breaking potential of the form as follows:
\be
{\cal L}_{dw} = - {1 \over 2} \na_{\mu} \varphi \na^{\mu} \varphi
- U(\varphi).
\ee
The symmetry breaking potential $U(\varphi)$ has a discrete set of
degenerate minima.  The energy-momentum tensor for the domain wall
yields
\be
T_{ij}(\varphi) = - {1 \over 2} g_{ij} \na_{m} \varphi \na^{m} \varphi
- U(\varphi) g_{ij} + \na_{i} \varphi \na_{j} \varphi.
\label{ten}
\ee
For the convenience we scale out parameters via transformation $ X =
{\varphi / \eta}$ and $\ep = 8 \pi G \eta^2$.  The parameter $\ep$
represents the gravitational strength and is connected with the
gravitational interaction of the Higgs field.  Defining $V(X) =
{U(\varphi) \over V_{F}}$, where $V_{F} = \lambda \eta^4$ we arrive at
the following expression:
\be
8 \pi G {\cal L}_{dw} = - {\ep \over w^2} \bigg[
w^2 {\na_{\mu} X \na^{\mu} X \over 2} + V(X) \bigg],
\label{dw}
\ee
where $w = \sqrt{{\ep \over 8 \pi G V_{F}}}$ represents the inverse
mass of the scalar after symmetry breaking, which also characterize
the width of the wall defect within the theory under consideration.
Having in mind (\ref{dw}) the equations for $X$ field may be written
as follows:
\be
\na_{\mu} \na^{\mu} X - {1 \over w^2}{ \p V \over \p X} = 0.
\ee
In the background of the dilaton black hole spacetime the
equation of motion for the scalar field $X$ implies
\be
{1 \over r \left ( r - {Q^2 \over M} \right )}
\p_{r} \bigg[ \big ( r - {Q^2 \over M} \big) \big( r - 2 M \big)
\p_{r} X \bigg] + 
{1 \over r \left ( r - {Q^2 \over M} \right ) \sin \theta }
\p_{\theta} \bigg[ \sin \theta \p_{\theta} X \bigg] =
{1 \over w^2}
{ \p V \over \p X}.
\label{motion}
\ee
Having in mind relation (\ref{ten}) one can define the energy
density of scalar fields $\varphi$ in the form
\be
E = {T_{t}{}{}^{t} \over \la \eta^4} =
\bigg[
- {1 \over 2} \big( X_{,r} \big)^2 \bigg( 1 - {2M \over r} \bigg)
-{ 1 \over 2} \big( X_{,\theta} \big)^2
{1 \over r \big(r - {Q^2 \over M} \big)}
 \bigg] { w^2} - V(X).
\label{energy}
\ee
In our considerations we take into account
for two cases of potentials with a discrete set
of degenerate minima, namely

\noindent
-- the $\varphi^4$ potential
\be
U_1(\varphi) = {\lambda \over 4} (\varphi^2 - \eta^2)^2,
\label{phi4}
\ee
-- the sine-Gordon potential
\be
U_2(\varphi) = \lambda \eta^4 \left [ 1 + \cos(\varphi/\eta) \right ].
\label{sinG}
\ee


\section{Boundary conditions and numerical integration.}
\subsection{Boundary conditions for $\varphi^4$ potential}
Because of the fact that the charged dilaton black hole is asymptotically
flat, the asymptotic boundary solution of equation of motion
for potential (\ref{phi4}) is the solution of equation of motion in flat
spacetime, namely
\be
\varphi_1(z) = \eta \tanh ( \sqrt{\lambda/2} \eta z).
\label{solphi4}
\ee
In our units this gives
\be
X(r,\theta) = \tanh \bigg( {r \cos \theta \over \sqrt{2} w }\bigg )
\ee
In this case $V(X) = {1 \over 4} (X^2-1)^2$ and $\p V/\p X = X(X^2-1)$
so the equation of motion (\ref{motion}) takes the form
\be
{1 \over r \left ( r - {Q^2 \over M} \right )}
\p_{r} \bigg[ \big ( r - {Q^2 \over M} \big) \big( r - 2 M \big)
\p_{r} X \bigg] + 
{1 \over r \left ( r - {Q^2 \over M} \right ) \sin \theta }
\p_{\theta} \bigg[ \sin \theta \p_{\theta} X \bigg] - {1 \over w^2}
X(X^2-1) = 0.
\label{motphi4}
\ee
while the energy density is equal to the following expression:
\be
E = \bigg[
- {1 \over 2} \big( \p_r X \big)^2 \bigg( 1 - {2M \over r} \bigg)
-{ 1 \over 2} \big( \p_\theta X \big)^2
{1 \over r \big(r - {Q^2 \over M} \big)}
 \bigg] {w^2} - {1 \over 4} (X^2-1)^2.
\label{enphi4}
\ee
On the horizon relation (\ref{motphi4}) gives the boundary condition
\be
{1 \over 2 M} \p_r X  \big|_{r=2M} = {- 1 \over 2M \left (2M - {Q^2
\over M} \right ) \sin \theta} \p_\theta \left [ \sin \theta \p_\theta
X \right ] + {1 \over w^2}
X(X^2-1).
\label{bhor}
\ee
Because we consider here only the case when the core of the wall is
located in the equatorial plane $\theta = \pi/2$ of the black hole we
impose the Dirichlet boundary condition at the equatorial plane
\be
X \big|_{\theta = \pi/2} = 0.
\label{beq}
\ee
The regularity of the scalar field on the symmetric axis requires the
Neumann boundary condition on the z-axis:
\be
{\p X \over \p \theta} \bigg|_{\theta=0} = 0.
\label{bz}
\ee
Far from the black hole we want to obtain flat spacetime solution
(\ref{solphi4}). Because our computational grid is finite this
requires the following boundary condition to be imposed on the outer
boundary of the grid:
\be
X \big|_{r=r_{max}} = \tanh \bigg({ r_{max} \cos \theta \over \sqrt{2} w} \bigg).
\label{brmax}
\ee

\subsection{Boundary conditions for the sine-Gordon potential.}
For the sine-Gordon potential (\ref{sinG}) the flat spacetime solution
is given by
\be
\varphi_2(z) = \eta \left \{ 4 \arctan \left [ \exp (\sqrt{\lambda}
\eta z ) \right ] - \pi \right \}.
\label{sinGsol}
\ee
In our units this is equivalent to following:
\be
X(r,\theta) = 4 \arctan \bigg [ \exp \bigg({r \cos \theta \over w}\bigg) \bigg ] - \pi.
\label{XsinG}
\ee
In this case $V(X) = 1 + \cos(X)$ and $\p V / \p X = - \sin(X)$, and
thus the equation of motion (\ref{motion}) takes the form
\be
{1 \over r \left ( r - {Q^2 \over M} \right )}
\p_{r} \bigg[ \big ( r - {Q^2 \over M} \big) \big( r - 2 M \big)
\p_{r} X \bigg] + 
{1 \over r \left ( r - {Q^2 \over M} \right ) \sin \theta }
\p_{\theta} \bigg[ \sin \theta \p_{\theta} X \bigg] + 
{1 \over w^2} \sin(X) = 0,
\label{motsinG}
\ee
while the energy density is given by the relation
\be
E = \bigg[
- {1 \over 2} \big( \p_r X \big)^2 \bigg( 1 - {2M \over r} \bigg)
-{ 1 \over 2} \big( \p_\theta X \big)^2
{1 \over r \big(r - {Q^2 \over M} \big)}
 \bigg] {w^2} - 1 - \cos(X).
\label{ensinG}
\ee
On the horizon from Eq.(\ref{motsinG}) one has the boundary condition as follows:
\be
{1 \over 2 M} \p_r X  \big|_{r=2M} = {- 1 \over 2M \left (2M - {Q^2
\over M} \right ) \sin \theta} \p_\theta \left [ \sin \theta \p_\theta
X \right ] - {1 \over w^2} \sin(X).
\label{bhorsinG}
\ee
Of course, this must be accompanied by the Dirichlet boundary conditions at the equatorial
plane of the black hole (\ref{beq}), and the Neumann
boundary condition on the symmetry axis (\ref{bz}),
and at the outer edge of the grid
\be
X \big|_{r=r_{max}} = 4 \arctan \bigg[ \exp \bigg({r_{max}\cos \theta \over w}\bigg)
\bigg ] - \pi.
\ee
%
\subsection{Numerical integration.}
In order to solve numerically
Eq.(\ref{motion}) we used the numerical method previously
used in Refs.\cite{dil}.
Namely, we use overrelaxation method slightly modified to
handle boundary conditions on the black hole horizon. We solve the
equation of motion on uniformly spaced polar grid $(r_i,\theta_i)$
with boundaries at $r_{min} = 2M$, outer radius $r_{max} \gg 2M$
(usually we use $r_{max} = 20M$), and $\theta$ ranging from $0$ to
$\pi/2$. The rest of the solution is obtained from symmetry of the scalar
field -- $X(r,-\theta) = X(r,\theta)$ and $X(r,\theta>\pi/2) =
-X(r,\pi-\theta)$.

\begin{figure}
\begin{center}
\includegraphics[width=330pt]{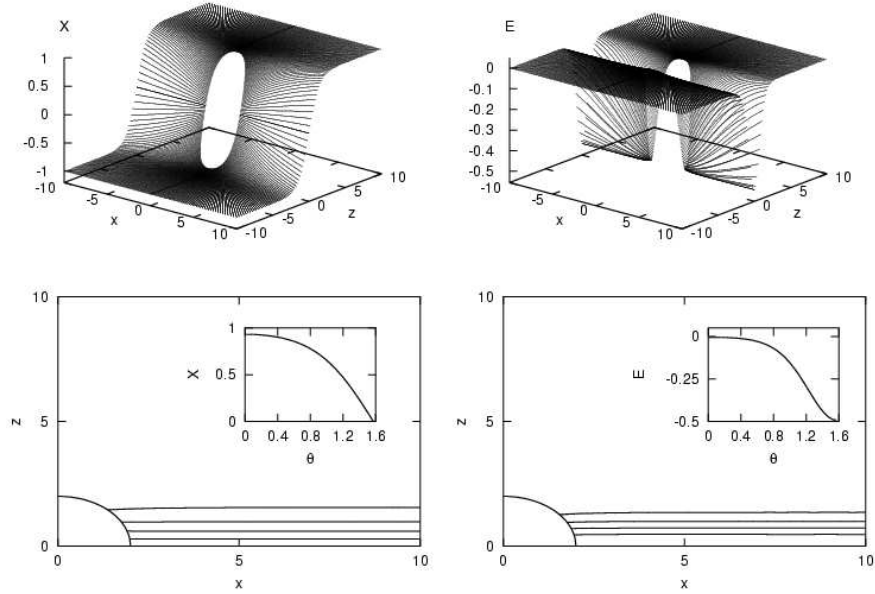}
\end{center}
\caption{
  The field X (left panels) and the energy E (right panels) for the
  $\phi^4$ potential. Isolines on bottom panels are drawn for $0.2$,
  $0.4$, $0.6$ and $0.8$ for the field X and for $-0.1$, $-0.2$,
  $-0.3$ and $-0.4$ for the energy. Inlets in bottom plots show the
  value of the fields on the black hole horizon. Black hole has
  $M=1.0$, $Q=0.1$ and the domain width is $w=1.0$.}
\label{figphi4}
\end{figure}

\begin{figure}
\begin{center}
\includegraphics[width=330pt]{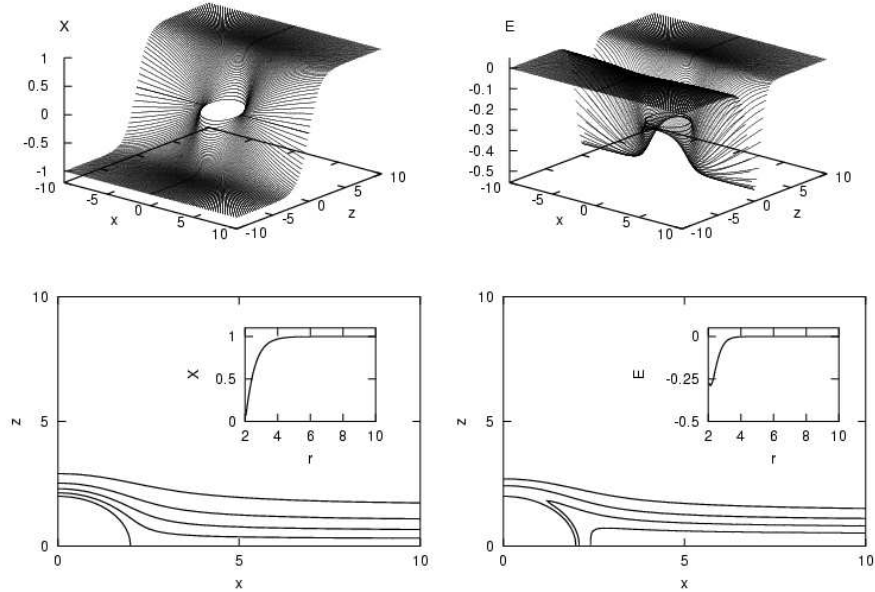}
\end{center}
\caption{
  The field X (left panels) and the energy E (right panels) for the
  $\phi^4$ potential and extreme black hole. Isolines on bottom panels
  are drawn for $0.2$, $0.4$, $0.6$ and $0.8$ for the field X and for
  $-0.1$, $-0.2$, $-0.3$ and $-0.4$ for the energy. Inlets in bottom
  plots show the value of the fields on the black hole z-axis. Black
  hole has $M=1.0$, $Q=\sqrt{2}$ and the domain width is $w=1.0$.}
\label{figphi4max}
\end{figure}

\begin{figure}
\begin{center}
\includegraphics[width=330pt]{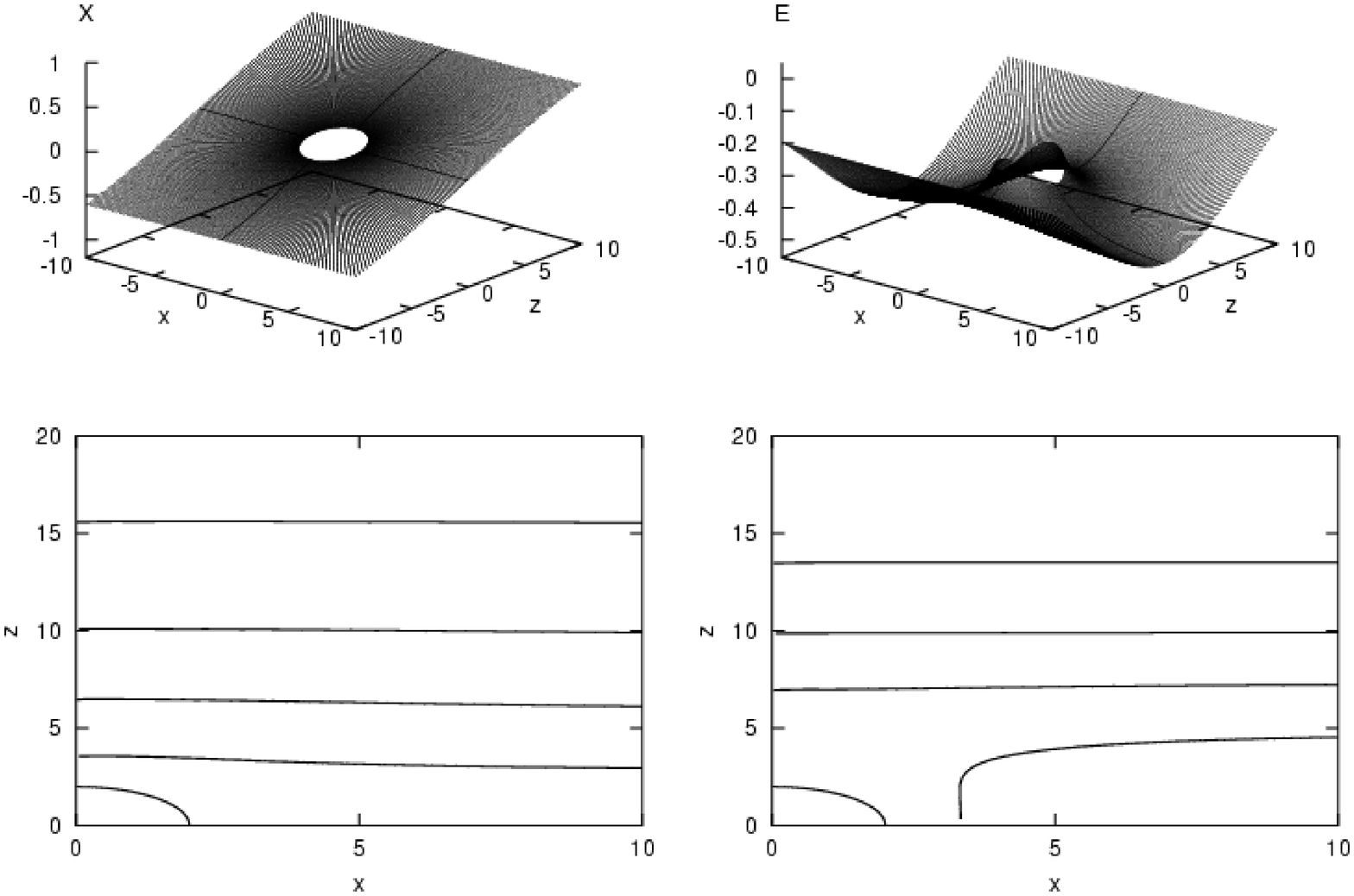}
\end{center}
\caption{
  The field X (left panels) and the energy E (right panels) for the
  $\phi^4$ potential. Isolines on bottom panels are drawn for $0.2$,
  $0.4$, $0.6$ and $0.8$ for the field X and for $-0.1$, $-0.2$,
  $-0.3$ and $-0.4$ for the energy. Black hole has
  $M=1.0$, $Q=0.1$ and the domain width is $w=10.0$.} 
\label{figphi410}
\end{figure}

\begin{figure}
\begin{center}
\includegraphics[width=330pt]{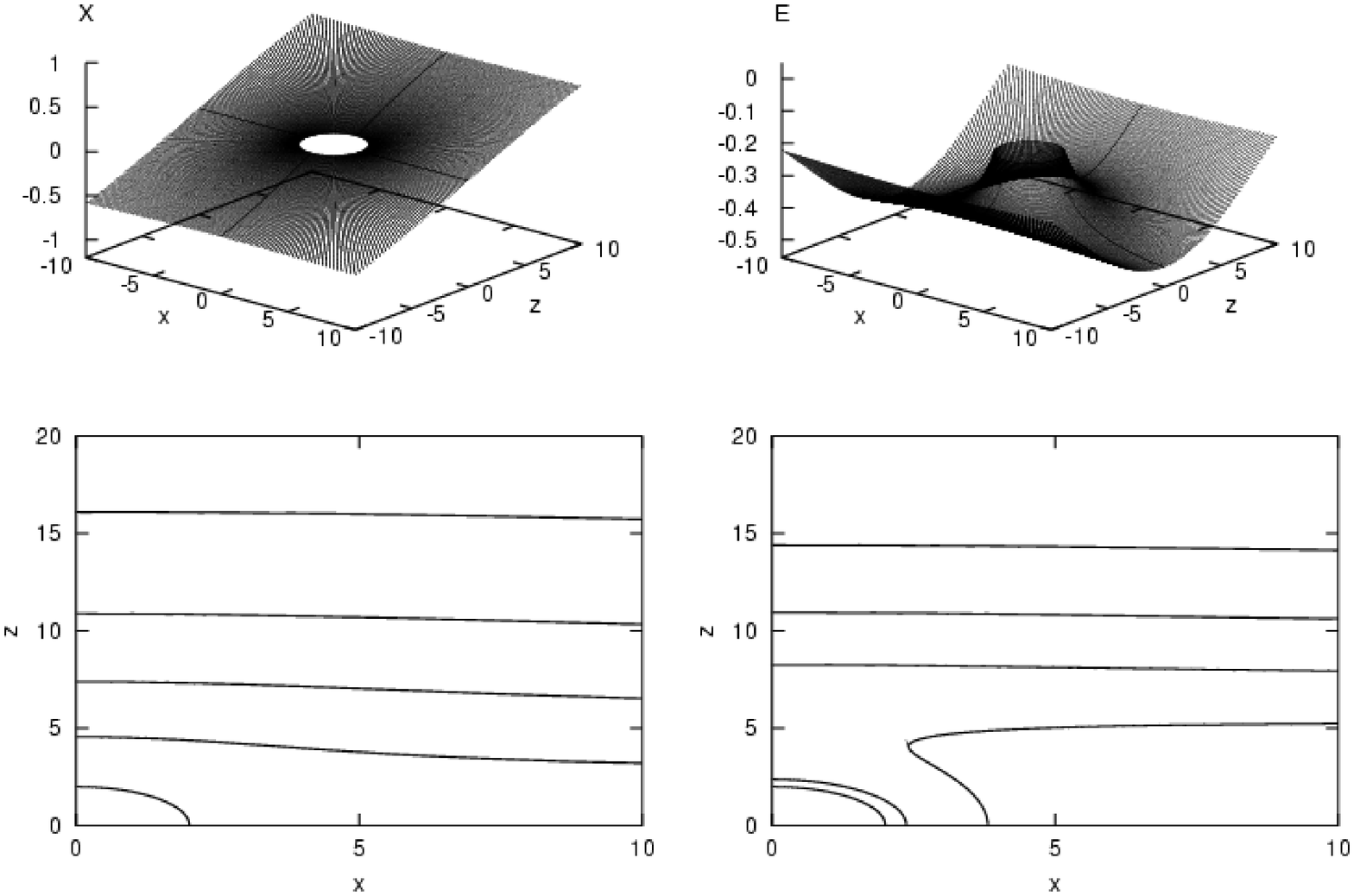}
\end{center}
\caption{
  The field X (left panels) and the energy E (right panels) for the
  $\phi^4$ potential and extreme black hole. Isolines on bottom panels
  are drawn for $0.2$, $0.4$, $0.6$ and $0.8$ for the field X and for
  $-0.1$, $-0.2$, $-0.3$ and $-0.4$ for the energy. The energy isoline
  around the black hole is $-0.3$. Black hole has $M=1.0$,
  $Q=\sqrt{2}$ and the domain width is $w=10.0$.}
\label{figphi410max}
\end{figure}

\begin{figure}
\begin{center}
\includegraphics[width=330pt]{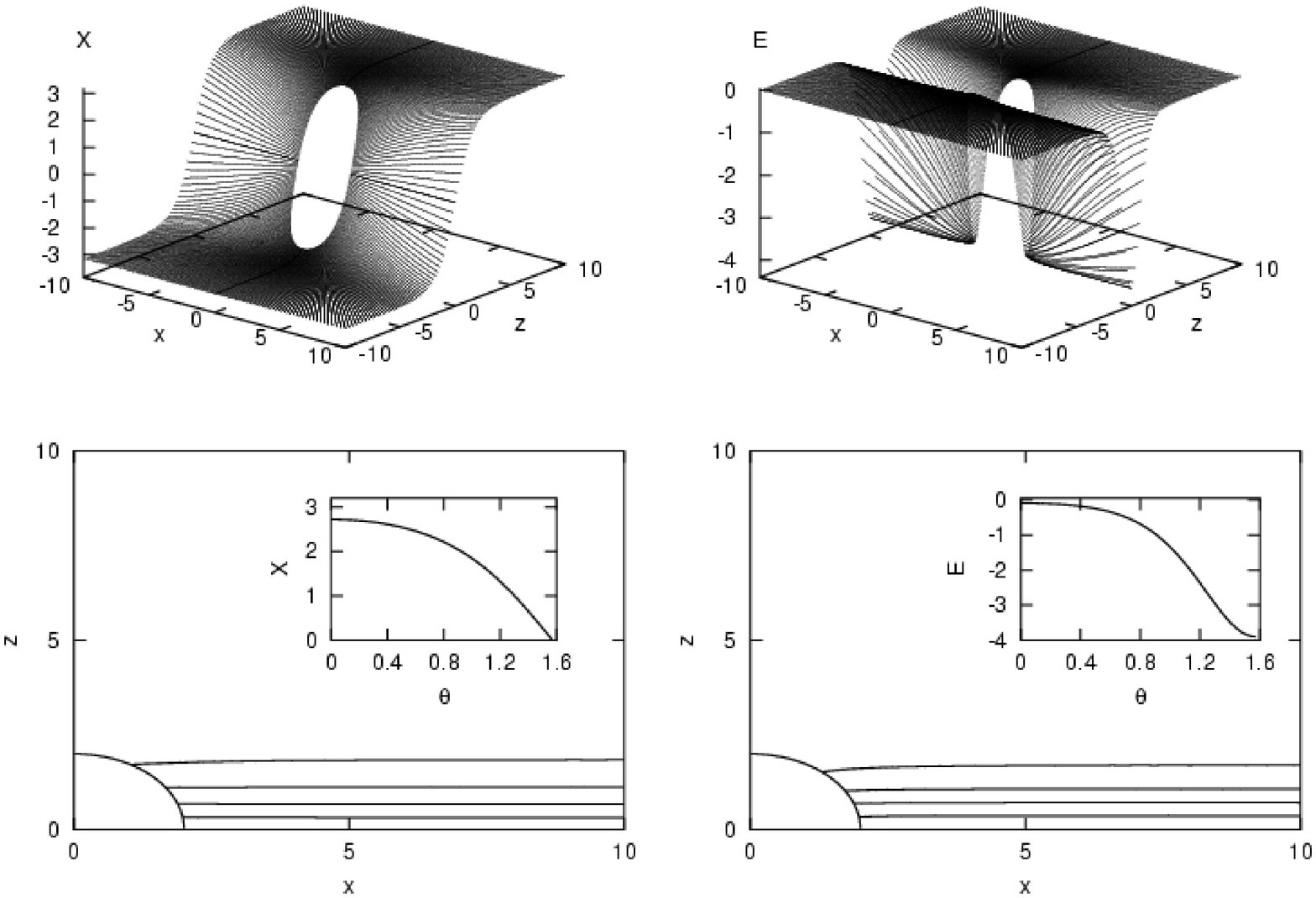}
\end{center}
\caption{
  The field X (left panels) and the energy E (right panels) for the
  sine-Gordon potential. Isolines on bottom panels are drawn for $0.2\pi$,
  $0.4\pi$, $0.6\pi$ and $0.8\pi$ for the field X and for $-0.5$, $-1.5$,
  $-2.5$ and $-3.5$ for the energy. Inlets in bottom plots show the
  value of the fields on the black hole horizon. Black hole has
  $M=1.0$, $Q=0.1$ and the domain width is $w=1.0$.}
\label{figsing}
\end{figure}

\begin{figure}
\begin{center}
\includegraphics[width=330pt]{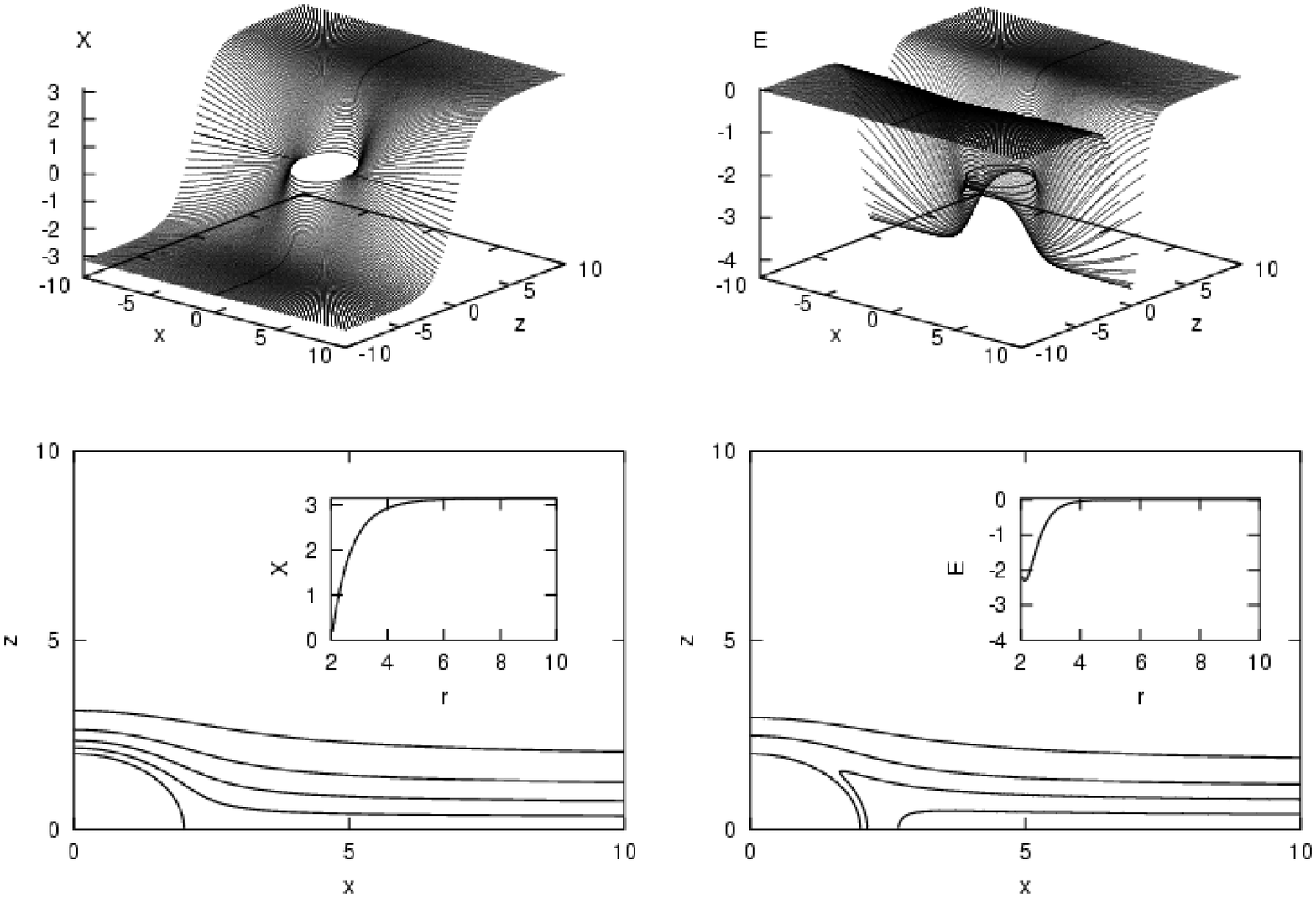}
\end{center}
\caption{
  The field X (left panels) and the energy E (right panels) for the
  sine-Gordon potential. Isolines on bottom panels are drawn for $0.2\pi$,
  $0.4\pi$, $0.6\pi$ and $0.8\pi$ for the field X and for $-0.5$, $-1.5$,
  $-2.5$ and $-3.5$ for the energy. Inlets in bottom plots show the
  value of the fields on the black hole z-axis. Black hole has
  $M=1.0$, $Q=\sqrt{2}$ and the domain width is $w=1.0$.}
\label{figsingmax}
\end{figure}

\begin{figure}
\begin{center}
\includegraphics[width=330pt]{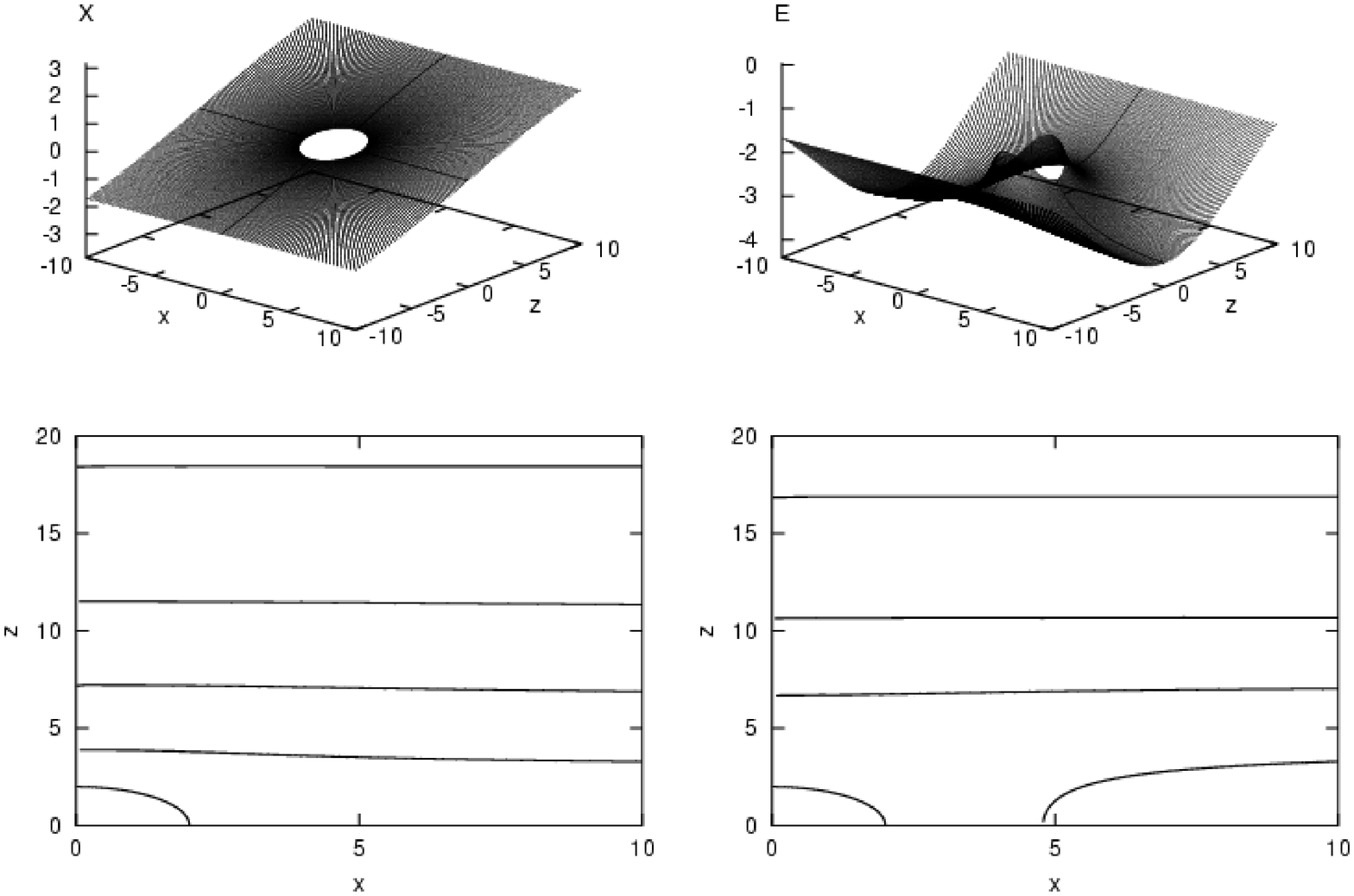}
\end{center}
\caption{
  The field X (left panels) and the energy E (right panels) for the
  sine-Gordon potential. Isolines on bottom panels are drawn for
  $0.2\pi$, $0.4\pi$, $0.6\pi$ and $0.8\pi$ for the field X and for
  $-0.5$, $-1.5$, $-2.5$ and $-3.5$ for the energy. Black hole has
  $M=1.0$, $Q=0.1$ and the domain width is $w=10.0$.}
\label{figsing10}
\end{figure}

\begin{figure}
\begin{center}
\includegraphics[width=330pt]{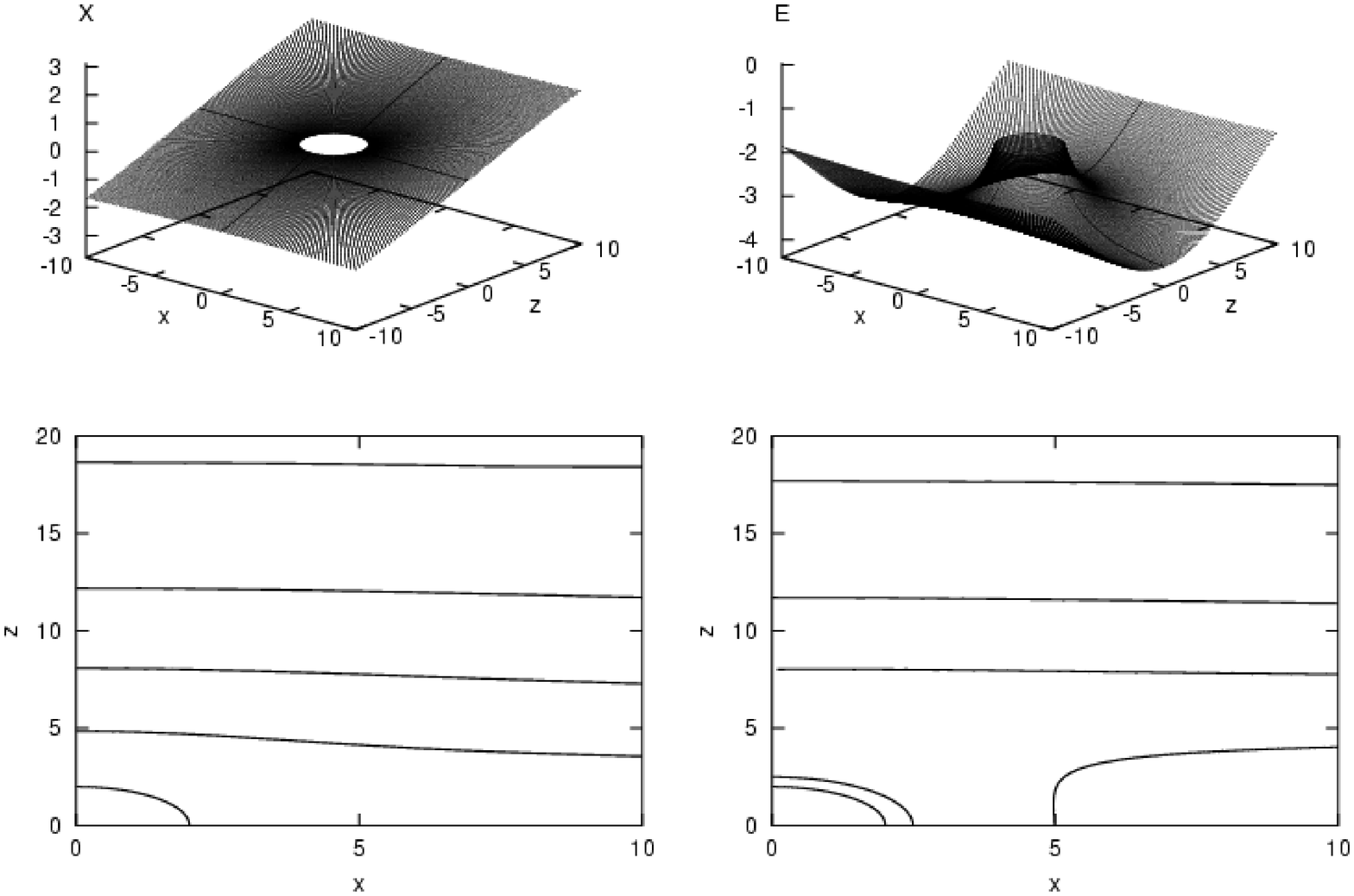}
\end{center}
\caption{
  The field X (left panels) and the energy E (right panels) for the
  sine-Gordon potential. Isolines on bottom panels are drawn for
  $0.2\pi$, $0.4\pi$, $0.6\pi$ and $0.8\pi$ for the field X and for
  $-0.5$, $-1.5$, $-2.5$ and $-3.5$ for the energy. The energy isoline
  around the black hole is $-2.5$. Black hole has $M=1.0$,
  $Q=\sqrt{2}$ and the domain width is $w=10.0$.}
\label{figsing10max}
\end{figure}

Figure~\ref{figphi4} presents the results of numerical integration of
equation of motion (\ref{motphi4}). On the same plot we show also the
energy (\ref{enphi4}) for this scalar field configuration. The mass of the
black hole is taken to be
$M=1$ and the charge $Q=0.1$, the domain wall thickness $w = 1$.
Fig.\ref{figphi4max} depicts the field $X$ and the energy $E$ for the extreme 
dilaton black hole with parameters $M = 1$, $Q = \sqrt{2}$ and the domain width
$w = 1$. For this case one has the expulsion of the domain wall from the
exteremal dilaton black hole. The so-called {\it Meissner effect} was analytically
predicted in Ref.\cite{rog01}.
Figs.\ref{figphi410} - \ref{figphi410max} were plotted for the same 
black hole parameters but we changed the domain wall width and put it
$w = 10$. Then for this width of the domain wall the dilaton black hole
is enveloped in the core region of the wall.
Figs.\ref{figsing} - \ref{figsingmax} show the values 
of $X$ field and the energy for the dilaton and extremal dilaton black holes
and the domain width $w = 1$. For this kind of potential one observes also the 
{\it Meissner effect} for the extreme black hole. 
In Figs.\ref{figsing10} - \ref{figsing10max} we take into account the domain width
$w = 10$. In this case we have also enveloping of the black hole in the core 
region of the considered domain wall.

\section{Conclusions}
In our paper we studied the problem of the domain wall in the vicinity
of a charged dilaton black hole, being the static spherically symmetric
solution of Einstein-dilaton gravity. We solve numerically Eqs. of motion
for real scalar field with $\varphi^4$ and sine-Gordon potentials. We use
the modified overrelation method modified to comprise the boundary 
conditions on the black hole event horizon. In our considerations we also use
the parameter $w = 1/\sqrt{\la} \eta$, which characterizes the thickness 
of the domain wall.
\par
We have justified the existence of static axisymmetric field configuration
representing thick domain wall in the nearby of dilaton black hole
for both $\varphi^4$ and sine-Gordon potential cases.
We studied the specific black hole domain wall configuration, 
when the core of
the domain wall is located at the equatorial plane of the black hole.
As in Ref. \cite{mor00} we assumed that the gravitational effect of the
domain wall is negligible compared to the effect caused by a
charged dilaton black hole.
In the case of the extreme dilaton black hole we find that the domain wall
is always expelled from the considered black hole, justifying the analytical
predictions presented in Ref. \cite{rog01}. This behaviour envisages the analog
of the so-called {\it Meissner effect} for the extreme dilaton black hole.
This effect was also revealed in the case of another topological defect,
i.e., cosmic string which was also expelled from the extremal dilaton
black hole \cite{dil,san00}.

%
\def\cmp#1#2#3{{ Commun. Math. Phys.} {\bf #1}, #2 (#3)}
\def\lmp#1#2#3{{ Lett. Math. Phys.} {\bf #1}, #2 (#3)}
\def\hpa#1#2#3{{ Hell. Phys. Acta} {\bf #1}, #2 (#3)}
\def\grg#1#2#3{{ Gen. Rel. Grav.} {\bf #1}, #2 (#3)}
\def\pr#1#2#3{{ Phys. Rev.} {\bf #1}, #2 (#3)}
\def\prl#1#2#3{{ Phys. Rev. Lett.} {\bf #1}, #2 (#3)}
\def\prd#1#2#3{{ Phys. Rev. D} {\bf #1}, #2 (#3)}
\def\pl#1#2#3{{ Phys. Lett} {\bf #1}, #2 (#3)}
\def\pla#1#2#3{{ Phys. Lett. A} {\bf #1}, #2 (#3)}
\def\plb#1#2#3{{ Phys. Lett. B} {\bf #1}, #2 (#3)}
\def\prep#1#2#3{{ Phys. Reports} {\bf #1}, #2 (#3)}
\def\phys#1#2#3{{ Physica} {\bf #1}, #2 (#3)}
\def\jcp#1#2#3{{ J. Comput. Phys.} {\bf #1}, #2 (#3)}
\def\jmp#1#2#3{{ J. Math. Phys.} {\bf #1}, #2 (#3)}
\def\jpm#1#2#3{{ J. Phys. A: Math. Gen.} {\bf #1}, #2 (#3)}
\def\cpr#1#2#3{{ Computer Phys. Rept.} {\bf #1}, #2 (#3)}
\def\cqg#1#2#3{{ Class. Quantum Grav.} {\bf #1}, #2 (#3)}
\def\cma#1#2#3{{ Computers Math. Applic.} {\bf #1}, #2 (#3)}
\def\mc#1#2#3{{ Math. Compt.} {\bf #1}, #2 (#3)}
\def\apj#1#2#3{{ Astrophys. J.} {\bf #1}, #2 (#3)}
\def\apjs#1#2#3{{ Astrophys. J. Suppl.} {\bf #1}, #2 (#3)}
\def\acta#1#2#3{{ Acta Astronomica} {\bf #1}, #2 (#3)}
\def\apl#1#2#3{{Ann. Physik. (Leipzig)} {\bf #1}, #2 (#3)}
\def\sa#1#2#3{{ Sov. Astro.} {\bf #1}, #2 (#3)}
\def\sia#1#2#3{{ SIAM J. Sci. Statist. Comput.} {\bf #1}, #2 (#3)}
\def\aa#1#2#3{{ Astron. Astrophys.} {\bf #1}, #2 (#3)}
\def\mnras#1#2#3{{ Mon. Not. R. astr. Soc.} {\bf #1}, #2 (#3)}
\def\npb#1#2#3{{ Nucl. Phys. B} {\bf #1}, #2 (#3)}
\def\prsla#1#2#3{{ Proc. R. Soc. London, Ser. A} {\bf #1}, #2 (#3)}
\def\jhep#1#2#3{{ JHEP} {\bf #1}, #2 (#3)}
\def\nuc#1#2#3{{Nuovo Cimento B } {\bf #1}, #2 (#3)}
\def\ijmp#1#2#3{{Int. J. Mod. Phys. D} {\bf #1}, #2 (#3)}

\def\hepth#1#2{{ hep-th }{\bf #1} (#2)}
\def\grqc#1#2{{ gr-qc }{\bf #1} (#2)}
%


\pagebreak

%
%
%
%
%
%
%
%


\end{document}